# Two SQUID amplifiers intended to alleviate the summing node inductance problem in multiplexed arrays of Transition Edge Sensors

**Mikko Kiviranta[1], Leif Grönberg[1] and Jan van der Kuur[2].**
[1] VTT Technical Research Centre of Finland, Tietotie 3, 02150 Espoo, Finland
[2] Space Research Organisation of the Netherlands, Landleven 12, 9747 AD Groningen, the Netherlands

E-mail: `Mikko.Kiviranta@vtt.fi`

**Abstract.** Frequency Domain Multiplexed detector arrays constructed of superconducting Transition Edge Sensors in the current-summing configuration suffer from the finite impedance of the summing node which should ideally be zero. We suggest two circuits to alleviate the effect. The first circuit uses a capacitive resonant transformer to increase the voltages and decrease the currents of TES signals to overcome the parasitic inductance of the interconnections. On the SQUID chip an impedance transform to the opposite direction takes place. The second circuit implements a power combiner having a better branch-to-branch isolation than a simple T-junction. Two SQUID devices have been designed and fabricated for a proof-of-principle demonstration of the circuits.

## 1. Introduction

Multiplexing [1] an array of Transition Edge Sensors (TESes) [2] involves fingerprinting the signals from individual TES pixels before they are summed into a so-called summing node. The total signal is then amplified, typically using a Superconducting Quantum Intereference Device (SQUID) [3] as the amplifier. The summing into a single node is occurs in Time Domain [4] and Code Domain [5] multiplexing schemes (TDM and CDM), too, but the summing is more problematic in the standard current-summing configuration of Frequency Domain multiplexing (FDM), where there is no SQUID-per-pixel active device providing the isolation between signals from different pixels. In the standard configuration, a significant constraint on the SQUID design is the required small input inductance, which (i) drives the SQUID energy resolution requirement, and (ii) limits the tolerable interconnection parasitics. We have experimented with two SQUID devices, first of which is targeted for the so called hi-Z summing node approach. The second device attempts to replace the T-junction signal combiner with a more complicated circuit providing isolation between the input branches. This work has been motivated by the X-IFU [6, 7] and SAFARI instruments [8] for space observatories on X-rays and submillimeter waves, respectively.

## 2. Summing node inductance

As recognized in the slides [1] and elaborated in [9], the common inductance is the impedance seen by the LC resonators as they feed the signals $I_1...I_N$ (fig. **1**a) to the summing node. Ideally the impedance should be zero when currents are summed, but in practice the inductance of interconnects and the SQUID input $L_{IN}$ (Fig. **1**) is significant enough to cause crosstalk between pixels. According to the X-IFU requirements the tolerated common mode inductance is order of 3 nH in the low-Z summing mode, comparable to the ~1 nH inductance due to a typical bonding wire. Already early FDM scenarios (slides [1], fig. 5 of [10]) made use of transformers close to TES pixels to up-transform the TES impedance, and to provide inductive rather than resistive biasing to the TESes. Another form of low-dissipating reactive TES bias using a capacitor [11] was suggested by van der Kuur, as well as use of the capacitive divider [12] as a resonant impedance transformer (fig. **1**b). The capacitive divider has the advantages that (i) in the case of up-transform the capacitor $C_{bN}$ (fig. **1**) defining the impedance ratio is physically smaller than the main resonant capacitor $C_{aN}$ whereas in the inductive case the bias coil would be physically larger; (ii) ratio of capacitances can be continuous while a the inductive trasformer typically has an integer turns ratio; and (iii) capacitive divider is better suited to fabrication capabilities at SRON who is responsible for the LC resonators within the

Two SQUID amplifiers intended to alleviate the summing node inductance problem in multiplexed arrays of Transition Edge Sensors

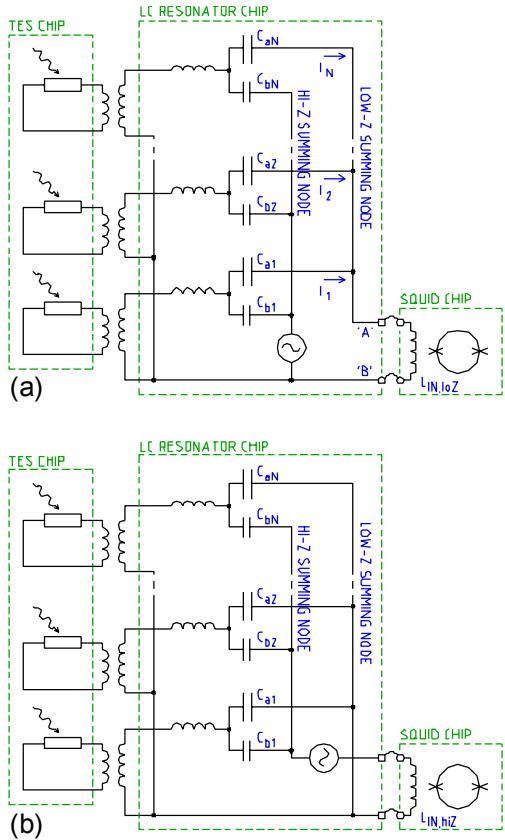

Figure 1: (a) A basic FDM system of fig. 2b with $C_{a1}$ … $C_{aN}$ as main resonating capacitors. Addition of small bias-injecting capacitors $C_{b1}$ … $C_{bN}$ allow main capacitors to double as TES bias sources. (b) When changing the roles of the two summing nodes, the capacitive Ca/Cb system acts a resonant transformer, called an L-section in the radio engineering parlance.

X-IFU consortium. The physical size of the resonator circuit is important because the resonators consume major amount of space in the X-IFU focal plane and we have already optimized the resonator size via the choice of impedance and hence the L-to-C ratio [13].

The high-Z summing point approach tolerates higher common inductance but requires a SQUID with higher current sensitivity. The energy resolution requirement $\varepsilon = \tfrac{1}{2} L_{IN} i_n^2$ would remain unchanged, expressed here neglecting back-action and in terms of input inductance and input-referred current noise. The device for the hi-Z summing node will be described in section 3.1 .

Because the low $L_{IN}$, essential to reach sufficient isolation between input branches, drives the $\varepsilon$ requirement in the circuits of fig. **1**, the question naturally arises whether the isolation could be arranged by using some other summing technique than the simple T-junction. Indeed, such signal power combiners are widely used at microwave frequencies, the best known probably being the Wilkinson combiner [14]. The Wilkinson is a λ/4 resonant version of the hybrid transformer [15, 16], which we have considered [17] either as cascaded 2-way combiners or as the generalized *N*-way combiner.

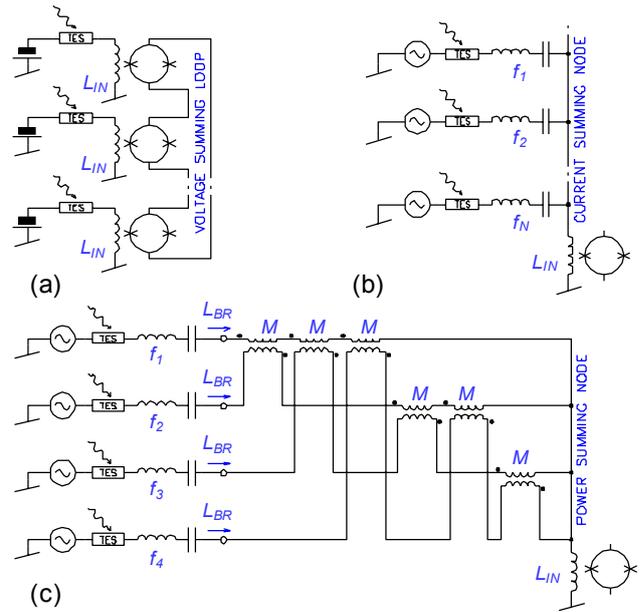

Figure 2: (a) In a TDM system even a large inductance $L_{IN}$ does not lead to crosstalk between channels as the inactive switch SQUIDs isolate TESes from the summing loop. (b) In an FDM system with T-junction current summing the voltage generated across the common inductance $L_{IN}$ leads to crosstalk. (c) When signal are summed by a generalized hybrid transformer circuit, with the choice of mutual inductances $M = L_{IN}$, the voltage across $L_{IN}$ due to an active branch gets cancelled in all other branches.

A 4-way version of the general *N*-way combiner is shown in fig. **2**c. The price to pay for the isolation between branches in that the inductance seen by each branch $L_{BR} = N\, L_{IN}$ is larger than in the T-junction circuit. This is an unavoidable consequence of the fact that all eigenvalues of the inductance matrix of a passively implementable coil system must be non-negative [18], the so-called Tokad-Reed condition. The branch inductance however can be absorbed as a part of the resonator inductance. Even though the *N*-way combiner with the full decoupling matrix would require $N\,(N-1)/2$ cancelling transformers, it may be enough to decouple just a few nearest-neighbour frequencies, as the stopband skirts of the LC resonators suppress crosstalk for signal pairs separated farther in frequency.

Two SQUID amplifiers intended to alleviate the summing node inductance problem in multiplexed arrays of Transition Edge Sensors

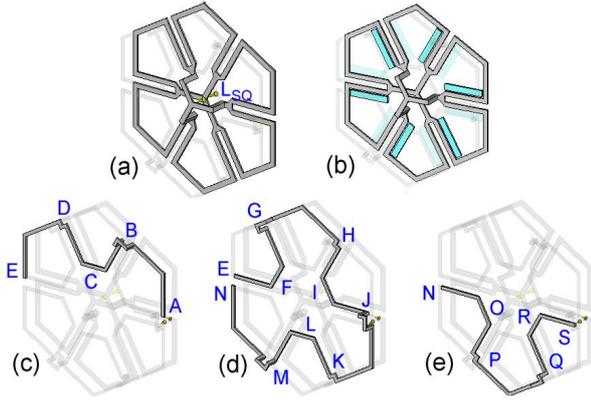

Figure 3: Out-of-scale presentation of the G1 core. (a) six niobium loops with alternating handedness, connected in parallel form the SQUID inductance $L_{SQ}$. (b) The junction shunt for one JJ is formed by six TiW resistors in parallel, each forming a transmission line with the adjacent niobium spoke. (c), (d) and (e) The niobium input coil traverses the path A-B-C-...-S in a higher wiring layer and forms the input inductance $L_{IN}$. The path is shown in three pieces for clarity.

The advantage of the hybrid transformer approach is that larger input inductance $L_{IN}$ of the SQUID can be tolerated, and energy resolution requirement is hence relaxed at a prescribed level of current noise $i_N$. A device implementing the hybrid transformer is shown in section 3.2 .

## 3. Device design and tests

Our designation for the hi-Z device is G2 and for the hybrid summing device G3. Both devices are build around the G1-type core SQUID, and come from the same mask set [17]. The G1 device [19, 20] is a 6-subloop counterwound fractional-turn SQUID [21]. Its magnetic coupling structure is schematically shown in fig. **3**. The 3 μm wide spokes are 200 μm long forming the $L_{SQ}$ = 70 pH SQUID inductance. Six parallel-connected resistors form a $R_S$ = 7.5 Ω shunt for the nominally 2.0 μm diameter Josephson junction, two of which are located at the center of the multiloop structure (nodes marked $L_{SQ}$ in fig. **3**a). The large volume of the resulting shunts improves electron-phonon coupling, in addition to which Au cooling fins are fabricated adjacent to the resistors. The magnetic field generated by the loops penetrates the cooling fins at signal frequencies, whereas eddy current losses contribute to microwave damping. The fabrication process has been described in [22].

*3.1 The G2 device for the hi-Z summing node*
In the G2 device the G1 core is equipped with an intermediate 12:1 fractional-turn transformer. This

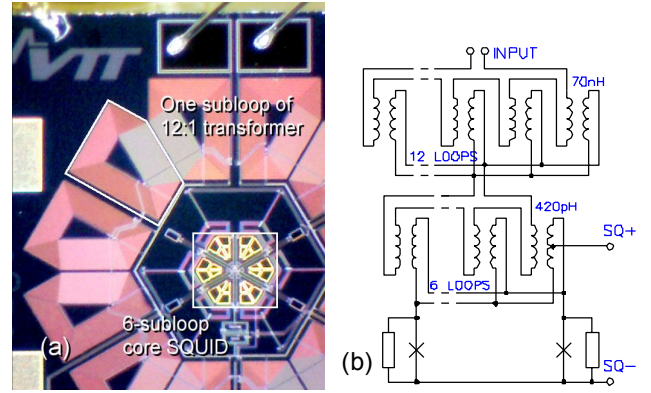

Figure 4: (a) A microphotograph of the G2-type SQUID device. (b) Magnetic coupling structure of the G2 consists of a 6-subloop fractional-turn SQUID inductance, combined with a 12-subloop fractional-turn intermediate transformer.

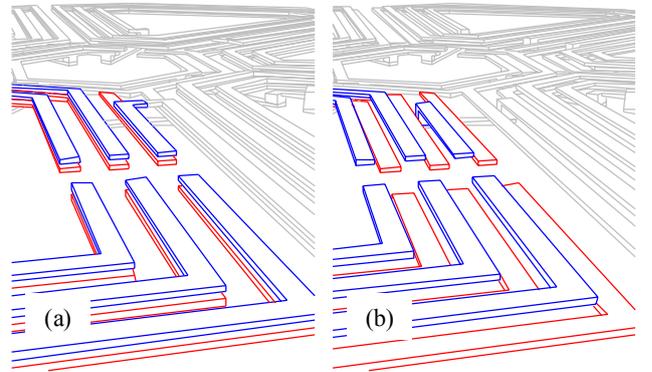

Figure 5: Cross-section of the schematical (a) doublestrip and (b) coplanar arrangement of one 1:1 subloop of an fractional-turn intermediate transformer. Blue is the upper NB2 metal containing the series-connected primary and red is the NB1 metal containing the parallel-connected secondary.

approach resembles [23] but has been conceived independently [17]. We also pursue the approach in an integrated SQUID magnetometer for combined MEG and MRI [24] where the spontaneous flux expulsion owing to narrow lines of the fractional-turn transformer is an advantage. The fig. **4**a is a microphotograph of the G2 device showing a part of the 2 x 2 mm chip. The fig. **4**b describes its magnetic configuration, where twelve 1:1 transformers are connected in series in the primary side, and in parallel in the secondary side. The G2 device exhibits $L_{IN} \approx 200$ nH input inductance, $M^{-1}$ = 0.8 μA/$\Phi_0$ input sensitivity, and $\Phi_N$ = 0.7 μ$\Phi_0$ / Hz$^{1/2}$ flux noise at $T$ = 4.2 K on selected operating points. Variation of the flux noise level depending on the operating point indicates that the microwave resonances [25] internal to the SQUID are not completely under control. A G2OF device variant

Two SQUID amplifiers intended to alleviate the summing node inductance problem in multiplexed arrays of Transition Edge Sensors exists, in which the transformer is split into the $M^{-1}$ = 1.2 µA/$\Phi_0$ input coil and $M^{-1}$ = 2.4 µA/$\Phi_0$ feedback coil for local linearization [26, 27].

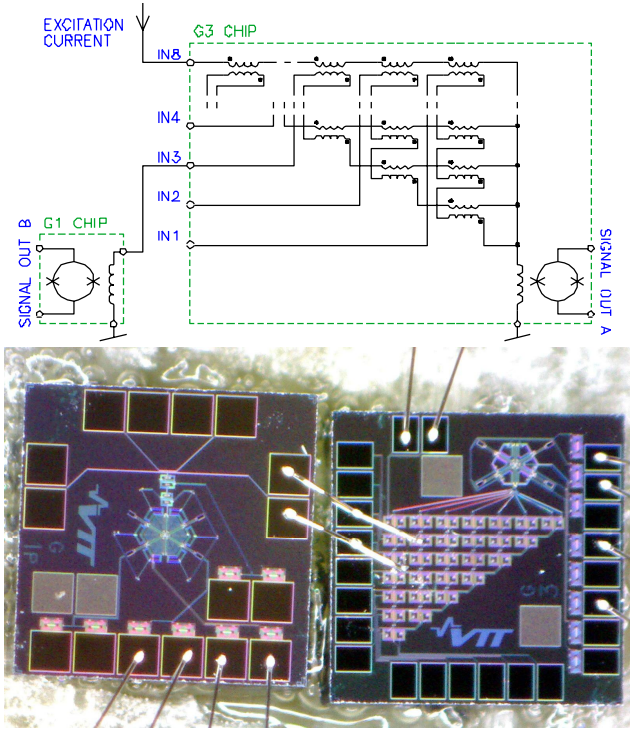

Figure 6: The experimental setup to test the power combiner: (top) schematic and (bottom) physical realization.

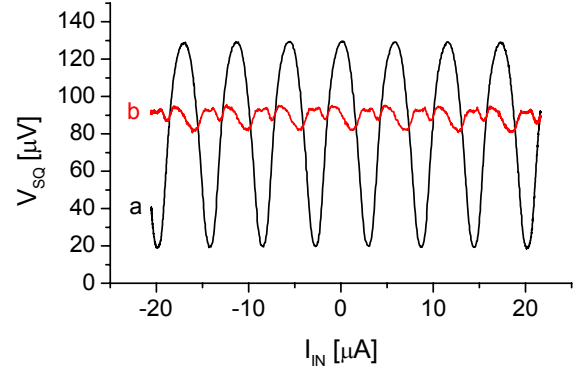

Figure 7: Flux-to-voltage response of the G3 main SQUID amplifier (a) and flux signal from the G1 used to observe the stray coupling (b).

Each subloop of the 12:1 intermediate transformer contains 16 turns, microwave damping problem [25] of which is beyond this paper. In short, we are making a tradeoff [28] between magnetic coupling and the Johnson noise due to the damping resistor. For this purpose we're employing a coplanar primary-secondary arrangement rather than the tighter coupled doublestrip structure (fig. **5**). With coplanar transformer magnetic coupling is looser, but higher-value microwave damping resistors could be used, generating less Johnson current noise. An experimental device variant with a doublestrip transformer was fabricated, and was observed to be unstable, as expected.

*3.2 The G3 device implementing the hybrid-based power combiner*

We have integrated a 8-way signal combiner with a G1-style 6-subloop SQUID core on a 2 x 2 mm chip (fig. **6**). We designate the device as G3. Although it would be more practical in an X-IFU -like FDM focal plane [7] to build the combiner on the LC resonator chip rather than SQUID chip, in order to minimize the number of chip-to-chip interconnects, our device acts as a demonstration of the concept. To verify the accuracy of the cancelling-transformer dimensioning and input-to-input isolation, we connected a G1-type SQUID chip to an inactive input of the G3 chip and drove excitation current to another input. The main signal from the G3 (fig. **7**a) is as expected. We observed input-to-input stray coupling to depend on the flux setpoint in the manner shown in fig. **7**b. We interpret this to be caused by variation of the input coil inductance of the SQUID core onboard the G3, due to the screening currents in the SQUID loop [29].

To assess the quality of input-to-input isolation, consider the $N \times N$ inductance matrices of an $N$-port combiner

$$\begin{bmatrix} L_{CM} & L_{CM} & \cdots & L_{CM} \\ L_{CM} & L_{CM} & & \\ \vdots & & \ddots & \\ L_{CM} & & & L_{CM} \end{bmatrix} \Rightarrow \begin{bmatrix} NL_{CM} & \alpha L_{CM} & \cdots & \alpha L_{CM} \\ \alpha L_{CM} & NL_{CM} & & \\ \vdots & & \ddots & \\ \alpha L_{CM} & & & NL_{CM} \end{bmatrix}$$

where the $i,j$:th entry corresponds to the voltage at the $i$:th port when current is injected to the $j$:th port at frequency $\omega$. The left case describes the $N$-way T-junction feeding the common inductance $L_{CM}$. The right case describes our hybrid-based combiner, where the parameter $\alpha$ is small, ideally $\alpha = 0$. From the signals of fig. **7** we estimate the realized $\alpha \approx 0.1$ in our case. We note that $\alpha$ is a function of flux setpoint, probably due to the screening currents within the SQUID core.

**4. Conclusion**

We have designed and fabricated two SQUID devices for two novel signal summing configurations, intended for Frequency Domain Multiplexed detector arrays. Proof-of-principle tests performed in liquid helium show that the devices function as designed. The G2 device having a large

Two SQUID amplifiers intended to alleviate the summing node inductance problem in multiplexed arrays of Transition Edge Sensors input inductance employs a novel fractional-turn intermediate transformer, and exhibits ε ≈ 300 h-bar energy resolution at $T = 4.2$ K. The G3 device implements a 8-way power combiner, which shows 90% reduction in common impedance at a cost of 8-fold increase in branch inductances, which however can be absorbed into LC resonator inductances. The remaining common impedance varies as a function of the flux setpoint, which we interpret as the screening due to the circulatory currents in the G3 SQUID loop. We foresee that the variation will limit the obtainable reduction in the common impedance.

**Acknowledgement**

This work has been supported by the European Union, from the E-SQUID project, grant no. 262947 of the 7[th] framework programme (FP7/2007-2013) as well as the AHEAD project, grant no. 654215 of the Horizon 2020 research and innovation programme. We are grateful for the assistance of Ms. Paula Holmlund and Mr. Harri Pohjonen for practical assistance.

**References**
[1] Kiviranta M, Seppä H, van der Kuur J and de Korte P 2001 SQUID-based readout schemes for microcalorimeter arrays, *9[th] International Workshop on Low Temperature Detectors (LTD-9),* published in *AIP Conf. Proc.* **605**, 295 (2002). Presentation slides online http://www.iki.fi/msk/xeus/ltd9slides.pdf .
[2] Ullom J N and Bennett D A 2015 Review of superconducting transition-edge sensors for x-ray and gamma-ray spectroscopy, *SuST*, **28** 084003 (36pp).
[3] Clarke J and Braginski A I (editors) 2004 *The SQUID handbook*, Wiley-VCH Verlag GmbH.
[4] Irwin K D 2002 SQUID multiplexers for transition-edge sensors, *Physica C*, **368** 203-10.
[5] Irwin K D et. al. 2010 Code-division multiplexing of superconducting transition-edge sensor arrays, *SuST*, **23** 034004 (7pp).
[6] Ravera L et. al. 2014 The X-ray Integral Field Unit (X-IFU) for Athena, *Proc. SPIE*, **9144** 91442L.
[7] Jackson B D et al. 2016 The focal plane assembly for the Athena X-ray Integral Field Unit instrument, *Proc. SPIE* **9905** 99052I.
[8] Jackson B D et al. 2012 The SPICA-SAFARI detector system: TES detector arrays with frequency-division multiplexed SQUID readout, *IEEE Tran. THz Sci. Tech*. **2** 12-21.
[9] Van der Kuur J et al. 2016 Optimising the multiplex factor of the frequency domain multiplexed readout of the TES-based microcalorimeter imaging array for the X-IFU instrument on the Athena X-ray observatory *Proc. SPIE*, **9905** 99055R. *ArXiv*:1611.05268 .
[10] Kiviranta M, van der Kuur J, Seppä H and de Korte P 2002 SQUID multiplexers for transition-edge sensors *Proceedings of the far-IR, sub-mm & mm detector technology workshop,* online http://www.iki.fi/msk/xeus/mm02_505pre.pdf .
[11] De Korte P et al. 2006 Frequency-domain multiplexed readout of EURECA, *7[th] International Workshop on Low Temperature Electronics (WOLTE-7),* 137-44, European Space Agency publication WPP-264.
[12] Van der Kuur J et al. 2012 The SPICA-SAFARI TES bolometer readout: Developments towards a flight system, *J. Low Temp. Phys.* **167** 561-7.
[13] Van der Kuur J and Kiviranta M 2002 Electronic readout concepts, document *SRON-CIS-Rep-WP-511*, European Space Agency, unpublished.
[14] Pozar D M 1998 *Microwave engineering*, John Wiley & sons, New York.
[15] Sartori E F 1968 Hybrid transformers *IEEE Trans. Parts, Mater. Packag.* **PMP-4** 59-66. DOI:10.1109/TPMP.1968.1135893
[16] "Understanding power splitters", *Application note AN10-006*, Mini-Circuits.
[17] Kiviranta M 2013 SQUID design phase II *Technical note TN6v2, TRP5417 project,* European Space Agency, unpublished.
[18] Tokad Y and Reed M 1960 Criteria and tests for realizability of the inductance matrix *Trans. AIEE, Pt I, Communications and Electronics* **78** 924-6. DOI: 10.1109/TCE.1960.6368492
[19] Kiviranta M, Grönberg L, Beev N and van der Kuur J 2014 Some phenomena due to SQUID input properties hen local feedback is present, *J. Phys. Conf. Series* **507** 042017.
[20] Gottardi L et al. 2015 Nearly quantum limited to-stage SQUID amplifiers for the frequency domain multiplexing of TES based X-ray and infrared detectors *IEEE Trans. Appl. Supercond.* **25** 2100404.
[21] Kiviranta M, Grönberg L and Hassel J 2012 A multiloop SQUID and a SQUID array with 1-μm and submicrometer input coils *IEEE Trans. Appl. Supercond.* **22** 1600105.
[22] Kiviranta M et al. 2016 Multilayer fabrication process for Josephson junction circuits cross-compatible over to foundries *IEEE Trans. Appl. Supercond.* **26** 1100905.
[23] Mates J A B et al. 2014 An efficient superconducting transformer design for SQUID magnetometry *J. Low Temp. Phys.* **176** 483-9.

[24] Luomahaara J et al. 2018 Unshielded SQUID sensors for ultra-low-field magnetic resonance imaging *IEEE Trans. Appl. Supercond.* **28** 1600204.
[25] Seppä H and Ryhänen T 1987 Influence of the signal coil on dc-SQUID dynamics *IEEE Trans. Magn.* **23** 1083-6.
[26] Irwin K D and Huber M E 2001 SQUID operational amplifier *IEEE Trans. Appl. Supercond.* **11** 1265-70.
[27] Kiviranta M 2008 SQUID linearization by current-sampling feedback *SuST.* **21** 045009.
[28] Seppä H and Kiviranta M 1999 Field sensitivity limitation due to pick-up coil resonances, *Extended abstracts of the 7th International Superconductive Electronics Conference (ISEC'99), June 21-25 1999, Berkeley, CA USA.*
[29] Hilbert C and Clarke J 1985 Measurements of the dynamic input impedance of a dc-SQUID *J. Low Temp. Phys.* **61** 237-62.